\documentclass[longbibliography,floatfix,reprint,twocolumn,notitlepage,prb,superscriptaddress,showpacs]{revtex4-1}
\usepackage{hyperref}
\usepackage[usenames,dvipsnames]{color}
\usepackage{amsmath}
\usepackage[english]{babel}
\usepackage{amssymb}
\usepackage{graphicx}
\usepackage{float}
\usepackage{epstopdf}
\usepackage{comment}
\usepackage{soul}
\usepackage{marvosym}
\usepackage{mathrsfs}
\usepackage{amsfonts}
\usepackage{tabularx}
\usepackage{multirow}
\usepackage{makecell}
\usepackage[normalem]{ulem}
\usepackage{wasysym}
\usepackage{dcolumn}
\newcolumntype{d}[1]{D{.}{.}{#1}}

\let\oldbibitem\bibitem
\renewcommand{\bibitem}{%
  \renewcommand{\doi}[1]{doi: ##1}
  \let\bibitem\oldbibitem
  \oldbibitem
}

\begin{document}

\title{Visible and ultraviolet plasma lines of the He-Ne gas laser}
\author{B. D. E. McNiven}
\email{bm2570@mun.ca}
\author{M. J. Clouter}
\author{G. Todd Andrews}

\affiliation{Department of Physics and Physical Oceanography, Memorial University of Newfoundland and Labrador, St. John's, Newfoundland \& Labrador, Canada A1B 3X7}

\date{\today}
\begin{abstract}
A study of Helium-Neon laser plasma lines was done using a double grating spectrometer and a He-Ne laser with an emission wavelength of 632.8 nm (15,802 cm$^{-1}$).  The absolute wavenumber, measured to within $\sim0.1$ cm$^{-1}$, and wavelength of each plasma line are presented, along with intensity and shift relative to the main laser line.   Several of the measured lines have not been reported in the literature and are observed at shifts between $0-1500$ cm$^{-1}$ from the laser line, a spectral region commonly probed by optical Raman scattering experiments. Accounting for the possibility of second-order diffraction permitted many previously unassigned lines to be attributed to known He or Ne electronic transitions with wavelengths in the ultraviolet region of the electromagnetic spectrum.
\end{abstract}

\maketitle
\section{Introduction}
Despite increasing adoption of solid state lasers, the He-Ne gas laser is still one of the most commonly used light sources in research and industrial applications.  Like all gas lasers, He-Ne lasers often emit so-called plasma lines arising from spontaneous emission in the laser tube in addition to the desired stimulated emission resulting from population inversion.  In laser spectroscopy experiments where precise knowledge of the wavenumbers associated with spectra features of interest are required, these plasma lines can serve as ``built-in'' calibration standards. It is therefore surprising that the literature on plasma emission lines for He-Ne lasers is scant. 

There appears to be only one in-depth study of He-Ne laser plasma emission lines \cite{Loader}. In this study, 38 plasma lines were observed over a spectral range of $12,093-15,798$ cm$^{-1}$.  The assignments of these lines and quoted wavenumbers, however, are questionable because the same study also includes results for Ar$^+$ laser plasma lines which have since come under scrutiny due to the apparent misidentification of several spectral peaks as plasma emission lines and to uncertainty in measured wavenumbers resulting from use of a spectrometer with relatively low resolution \cite{Craig}.  There are also differences in He-Ne plasma line wavenumber assignments as reported by Ref. \cite{Loader} and those in a Raman scattering study on multiferroic compounds. In particular, Raman spectra of CuFeO$_2$ and CuCrO$_2$ yielded strong plasma lines in the vicinity of 137 cm$^{-1}$, 180 cm$^{-1}$, and 200 cm$^{-1}$ \cite{Aktas_2012}, the former two differing by $\geq2$ cm$^{-1}$ when compared to the previous data \cite{Loader}, and the latter one being completely absent.

Motivated by the above-noted paucity of literature data and the inconsistencies in that which is available, we present in this paper the results of a comprehensive study on plasma lines from a He-Ne laser with an emission wavelength of 632.8 nm.  We report plasma lines with shifts of $0-1500$ cm$^{-1}$ from the primary emission line at 15802.4 cm$^{-1}$, a commonly probed spectral range for optical Raman scattering studies, and tabulate absolute wavenumber, measured intensity, and wavenumber shift from the primary emission line.  This information is also provided for some second-order lines in the ultraviolet spectral region. In addition, the atomic species (He or Ne) associated with each transition is identified.

\section{Experimental Details} \label{sec:exp_details}
\begin{figure}[!htb]
    \centering
    \includegraphics[width=0.3\textwidth]{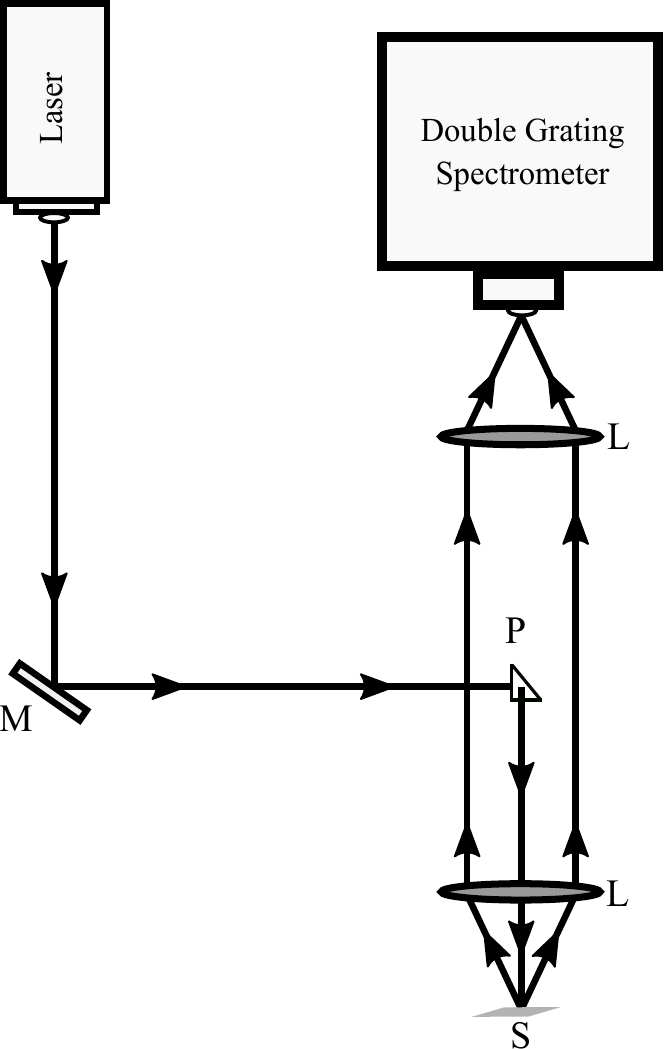}\\
    \caption{Experimental set-up.  M - front surface mirror, P - prism, L - lens, S - sample.}
    \label{fig:setup}
\end{figure}

Fig. \ref{fig:setup} shows a schematic diagram of the apparatus used to collect the emission spectra. The experiments were done in air at room-temperature   with a He-Ne laser emitting at a wavelength of 632.8 nm. The incident power was $\leq$5 mW, making observation of unwanted peaks due to inelastic light scattering much less likely. The beam was focused onto one of the three samples using a $f=5$ cm lens.  The reflected/scattered light was collected by the same lens ({\it{i.e.}}, backscattering configuration) and focused onto the entrance slit of a Spex model 1401 double grating spectrometer using a lens of focal length $f=25$ cm. All three spectrometer slits (entrance, center, and exit) were set to a width of 350 $\mu$m. The estimated spectral resolution is $\sim3$ cm$^{-1}$. The light exiting the spectrometer was detected by a low dark count rate ($\sim1$ s$^{-1}$) photomultiplier tube.

Experiments were conducted over shifts of $0-1500$ cm$^{-1}$ relative to the laser wavelength at 632.8 nm with crystalline silicon, aluminum, and superconductor Bi$_2$Sr$_2$CaCu$_2$O$_{8.17}$ as scattering sources to ensure that the origin of peaks assigned to plasma lines was indeed emission from the laser tube and not inelastic scattering from a particular material.  To ensure proper calibration of the spectrometer before each run, the signal from the primary emission line was first obtained over the range 15,809.8 cm$^{-1}$ to 15,800.5 cm$^{-1}$ with a grating step size of 0.05 cm$^{-1}$ by placing several neutral density filters in the beam path to heavily attenuate the incident light.  Following this, the filters were removed and spectrum collection was resumed (with no change in scan direction so as to eliminate possible dial backlash issues) at a starting wavenumber of 15,800.5 cm$^{-1}$ and a step size of either 0.15 or 0.3 cm$^{-1}$, corresponding to total acquisition times of $\sim222$ hours and $\sim111$ hours, respectively.
\section{Results and Discussion}

\begin{figure*}[htb]
\hspace*{-2cm}
\vspace*{-1cm}
\centering
\includegraphics[width=1.2\textwidth]{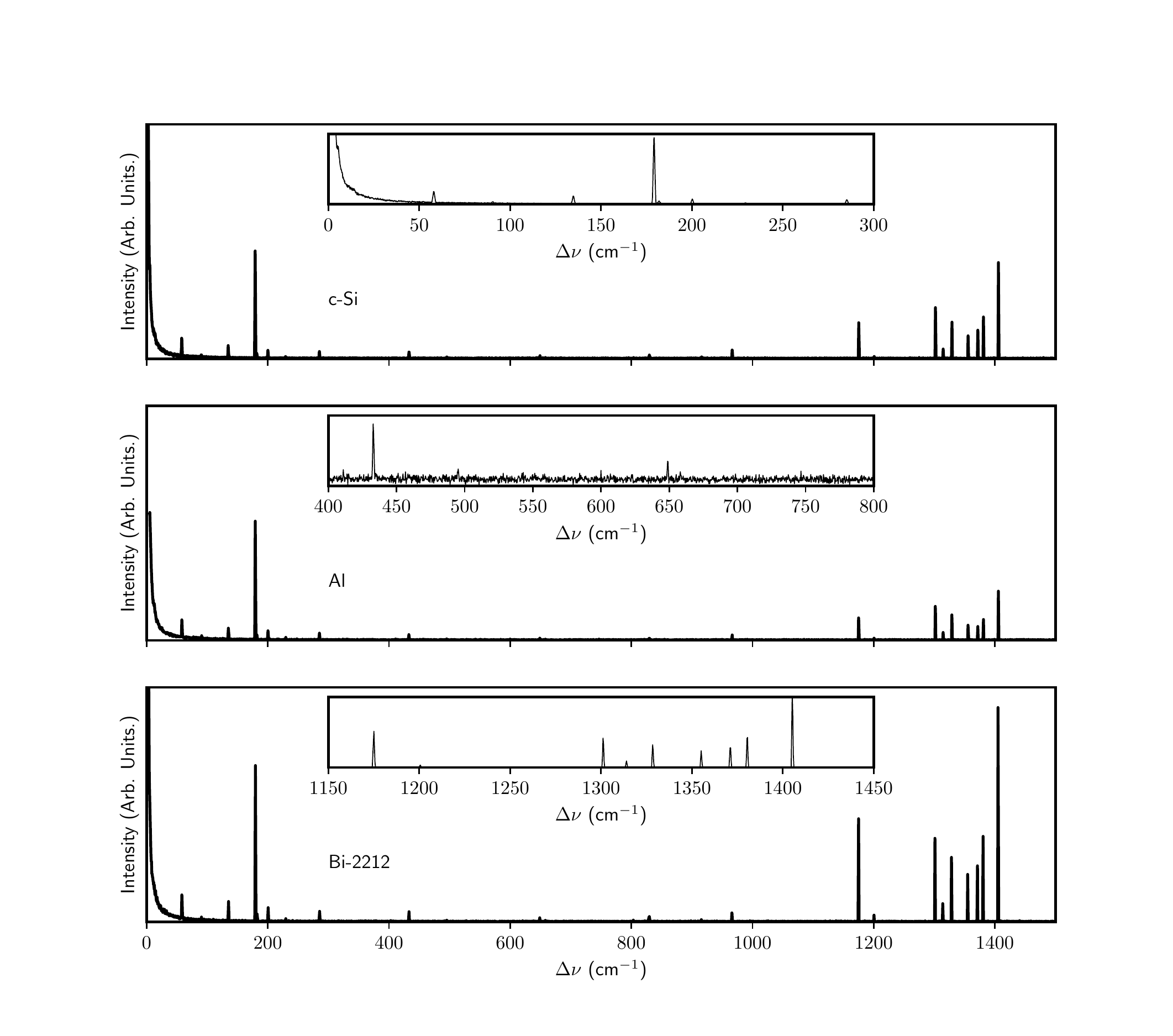}\\
    \caption{Emission spectra of a He-Ne laser with primary emission wavelength of 632.8 nm obtained with three different scattering sources.  Multiple plasma lines are present at the same shifts in all three spectra. Top Panel - crystalline silicon (c-Si); Middle Panel - aluminum (Al); Bottom Panel - Bi-2212 - Bi$_2$Sr$_2$CaCu$_2$O$_{8.17}$. Insets show zoomed in spectral regions of particular interest.}
    \label{fig:Raman}
\end{figure*}

Fig. \ref{fig:Raman} shows spectra collected in air of laser light scattered from each of the three materials. The spectra are very similar to one another, suggesting that our assignments of multiply-observed peaks to plasma emission lines are robust.  In order to be classified as a plasma line, it was required that the line be present in spectra of all three materials at the same shift.  The peak positions reported in this paper are those obtained for c-Si as the scattering material because its spectrum was collected for the longest time with the smallest step size.  As seen in the top and bottom insets of Fig. \ref{fig:Raman}, the spectral regions between $0-300$ cm$^{-1}$ and $1200-1500$ cm$^{-1}$ contain particularly large numbers of plasma lines.  In contrast, the center inset of Fig. \ref{fig:Raman} shows that relatively few lines are present in the spectral region from 400 cm$^{-1}$ to 800 cm$^{-1}$. Spectral peak parameters were obtained by fitting a Voigt function to each peak using the ``Fityk'' routine \cite{fityk}.  All lines were found to have a width of $\sim0.3$ cm$^{-1}$, suggesting a common origin, thereby providing further support for their assignment as plasma emission lines.

Comparison of our spectra to the data in Ref. \cite{Loader} reveals multiple plasma peaks that were not observed in the latter study (the criterion for being labelled as a `different' line being that the shift differed by more than $\sim1$ cm$^{-1}$ from the closest line in our spectra).  Our spectra in Fig. \ref{fig:Raman} also confirm the presence of a presumed plasma line at a shift of $\sim200$ cm$^{-1}$ reported in a previous Raman scattering study \cite{Aktas_2012}, although we were unable to assign this line to a particular He or Ne electronic transition.  This line is also missing from the earlier compilation \cite{Loader}. 

The absence in the literature of multiple lines that appear in our spectra is surprising, especially given that our results indicate that the intensity of many of the these lines is greater than that of some previously observed \cite{Loader}.  This can be largely reconciled by noting that ultraviolet lines corresponding to known He or Ne transitions with emission wavelengths of $\sim300$ nm \cite{fuhr2005nist} can satisfy the grating equation for second-order diffraction.  Furthermore, there are several lines listed in Ref. \cite{Loader} (642.17108 nm, 644.47118 nm, and 666.68967 nm, identified by `$\dagger'$ in Table \ref{tab:plasmalines}) that are present in our spectra but do not correspond to known He or Ne transitions in the visible region of the spectrum.  If, however, these lines are instead assumed to be second-order ultraviolet lines with wavelengths of $\sim300$ nm then they closely match our lines and can be associated with known He or Ne transitions.  With these points in mind, we present in Table \ref{tab:plasmalines} the wavelength in air ($\lambda_a$) and vacuum ($\lambda_v$), absolute wavenumber in air ($\nu_a$) and vacuum ($\nu_v$), shift relative to the 632.8 nm primary emission line ($\Delta\nu$), and intensity of each of the observed plasma emission lines shown in Fig. \ref{fig:Raman}.  We also list vacuum wavenumbers of He or Ne electronic transitions likely responsible for the observed lines.  Analogous data from Ref. \cite{Loader} is also included for the purposes of comparison. 

\begin{table*}[t]
\caption{Wavelength in air ($\lambda_a$) and vacuum ($\lambda_v$), absolute wavenumber in air ($\nu_a$) and vacuum ($\nu_v$), wavenumber shift relative to 15,802.4 cm$^{-1}$ in air ($\Delta\nu_a$), and intensity of plasma lines for a He-Ne laser emitting at 632.8 nm. $m$ - order of diffraction, PW - present work. The refractive index of air used to convert $\lambda_a$ to $\lambda_v$ was 1.00027653 for $m=1$ and 1.00029 for $m=2$ \cite{Ciddor96}. The listed uncertainties apply only to the present work and were obtained by collecting multiple spectra over a specific spectral range and calculating the standard deviation from resulting best-fit peak parameters. A `$\dagger$' by data from Ref. \cite{Loader} indicates possible $m=2$ lines.}
\begin{ruledtabular}
\begin{tabular}{c|ccccccc|c}
\multirow{2}{*}{Study} & $\lambda_{v}$ & $\nu_v$ & $\lambda_{a}$ & $\nu_a$ & $\Delta \nu_a$ & Intensity & \multirow{1}{*}{Diff in $\nu_v$}&\multirow{2}{*}{\thead{Elemental \\ Trans ($\nu$ cm$^{-1}$)}} \\ 
 & ($\pm$0.004 nm) & ($\pm$0.1 cm$^{-1}$) & ($\pm$0.004 nm) & ($\pm$0.1 cm$^{-1}$) & ($\pm$0.1 cm$^{-1}$) & ($\pm10\%$ A.U.) & ($\pm$0.1 cm$^{-1}$) & \\ \hline 
\multirow{13}{*}{\makecell{PW(m=1)}} & 632.991 & 15798.0 & 632.816& 15802.4 & 0 & -- & -- & --  \\
& 633.188 & 15793.1 &  633.013 & 15797.5 & 4.9 & 1421  & -- & Ne (15795.5)\footnotemark[1] \\
& 635.318 & 15740.1 &  635.142 & 15744.5 & 57.9  & 305  & +1.0 & Ne (15743.3)\footnotemark[1] \\ 
& 638.431 & 15663.4 & 638.255 & 15667.7 & 134.7 & 199  & +1.0 & Ne (15666.6)\footnotemark[1]\\
& 640.252 & 15618.9 & 640.075 & 15623.2   & 179.2   &  1611  & -- & Ne (15622.3)\footnotemark[1] \\
& 640.363 & 15616.1 & 640.186  & 15620.5 & 181.9 & 85  & +0.9 & Ne (15619.6)\footnotemark[1] \\
& 650.825 & 15365.1 &  650.645 & 15369.4 & 433.0 & 103  & +0.2 & Ne (15369.2)\footnotemark[1]\\
& 653.485 & 15302.6 &  653.304 & 15306.8 & 495.6 & 30 & -0.4 & Ne (15307.1)\footnotemark[1]\\
& 660.115 & 15148.9 & 659.933 & 15153.1 & 649.3 & 49  & -0.8 & Ne (15153.8)\footnotemark[1] \\
& 660.516 & 15139.7 &  660.333 & 15143.9   & 658.5   & 22  & -- & Ne (15144.9)\footnotemark[1] \\
& 668.071 & 14968.5 &  667.886 & 14972.6 & 829.8 & 59  & -1.3 & He (14970.8)\footnotemark[2] \\
& 671.942 & 14882.2 & 671.756 & 14886.4 & 916.0 & 31 & -1.2 & Ne (14887.6)\footnotemark[1]\\
& 693.168 & 14426.5 & 692.976 & 14430.5 & 1371.9 & 428  & -0.6 & Ne (14431.1) \footnotemark[1] \\
\hline
\multirow{11}{*}{\makecell{PW(m=2)}}&  318.317  & 31415.2 & 318.225 &  31424.3    & 90.2     &  60     & -- & He (31385.2)\footnotemark[1]\\ 
& 321.161 & 31137.0 & 321.068 & 31146.1 & 229.3 & 36 & -- & He (31128.4)\footnotemark[2]\\
& 322.318 & 31025.2 & 322.225& 31034.2 & 285.3 & 111  & -- & Ne (31009.7)\footnotemark[1]\\
& 333.463 & 29988.3 & 333.366& 29997.1 & 803.8 & 22  & -- & Ne (29986.8)\footnotemark[3] \\
&  337.117 & 29663.3 & 337.019& 29671.9  & 966.4   & 133   &  -- & Ne (29657.1)\footnotemark[3]\\
& 341.932 & 29245.6 & 341.833& 29254.1   & 1175.4   & 541   & -- & Ne (29256.9)\footnotemark[1]\\
& 342.529 & 29194.6 & 342.430  & 29203.0   & 1200.9   & 37   & -- & Ne (29206.5)\footnotemark[1]\\
& 344.918 & 28992.4 & 344.818 & 29000.8 & 1302.0   & 766   &  -- & Ne  (29004.8)\footnotemark[1]\\
& 345.221 &  28967.0 & 345.121 & 28975.3   & 1314.7   & 148   &  -- & Ne (28959.5)\footnotemark[1]\\
& 345.568 & 28937.9 & 345.468 & 28946.2   & 1329.3   & 549   &  -- & He (28928.1)\footnotemark[4]\\
& 346.203 & 28884.8 & 346.103   & 28893.1  & 1355.8   & 345   &  -- & Ne (28875.9)\footnotemark[3]\\
& 346.813 & 28834.0 & 346.712  & 28842.4   & 1381.2   & 626   &  -- & Ne (28846.7)\footnotemark[1]\\
& 347.409 & 28784.5 & 347.308 & 28792.9   & 1405.9   & 1438   & -- & Ne (28775.3)\footnotemark[1]\\ \hline
\multirow{1}{*}{\makecell{PW(m=?)}} & \multirow{1}{*}{?} & \multirow{1}{*}{?} & \multirow{1}{*}{?} & \multirow{1}{*}{?}   & \multirow{1}{*}{200.2}   & \multirow{1}{*}{129}  &  \multirow{1}{*}{--} & \multirow{1}{*}{?} \\
\hline \hline
\multirow{15}{*}{\makecell{Ref.\cite{Loader}}} &  & 15798.002 & 632.81646& &  & &\\
& & 15782.381 &  633.44279&  &   & & \\
& & 15739.064 & 635.18618&  &  & &  & \\
&  & 15662.306 & 638.29914&  &  & & & \\
&  & 15615.202 &  640.22460 & & & &  &\\
& $\dagger$ & 15567.871 &  642.17108&  &  & &  &\\
& $\dagger$ &  15512.310 & 644.47118& &  & &  & \\
& & 15364.935 & 650.65279 &  &  & &  &\\
& & 15302.951 & 653.28824 & &  & &  &\\
& & 15149.735 &659.89529& &  & & & \\
&  & 15028.714 &665.20925& &  & & & \\
& $\dagger$ & 14995.342 & 666.68967& &  & & & \\
&  & 14969.790 & 667.82764&  &   & &  &\\
&  & 14883.395 & 671.70428& &  & &  &\\
&  & 14427.144 &692.94672& &   & &  &\\
\end{tabular}
\end{ruledtabular}
\label{tab:plasmalines}
\footnotetext[1]{Vacuum wavenumbers from Ref. \cite{fuhr2005nist}.}
\footnotetext[2]{Air wavenumbers from data compilation in Ref. \cite{Martin}.}
\footnotetext[3]{Vacuum wavenumbers from Ref. \cite{GovPub}.}
\footnotetext[4]{Vacuum wavenumbers from Ref. \cite{Herzberg}.}
\end{table*}

Table \ref{tab:plasmalines} reveals some notable differences between the current and previously published results.  Firstly, several of our measured plasma line wavenumbers differ from those of Ref. \cite{Loader} by $\gtrsim1$ cm$^{-1}$ (see second column from the right in Table \ref{tab:plasmalines}). Secondly, Ref. \cite{Loader} reports plasma peaks near shifts of 15.63 cm$^{-1}$ and 769.51 cm$^{-1}$. We also observe a feature at $\sim15$ cm$^{-1}$, but no peak is present in our spectra at the latter wavenumber shift.  While not well-resolved, the general shape (broad and weak) of the feature at $\sim15$ cm$^{-1}$ does not appear to be characteristic of a typical plasma line (see top inset in Fig. \ref{fig:Raman}).  We note, however, that there is a known Ne transition at $\lambda_v=633.44$ nm \cite{fuhr2005nist} corresponding to a shift in air of $\sim11$ cm$^{-1}$ from the primary emission line and therefore it is possible that the feature at $\sim15$ cm$^{-1}$ is in fact a plasma line.  The peak at a shift of $\sim$770 cm$^{-1}$ reported in Ref. \cite{Loader}, although absent in our spectra, also cannot be discounted as a possible plasma line because there is a known Ne transition with an emission wavelength of $\lambda_v=332.72$ nm corresponding to $\Delta \nu_a = 770.391$ cm$^{-1}$ \cite{fuhr2005nist}.

\section{Conclusions}
Plasma lines from a He-Ne laser source were investigated by probing a spectral region $0-1500$ cm$^{-1}$ from the primary emission wavelength of 632.8 nm ($\nu_a = 15,802.4$ cm$^{-1}$). From the collected spectra, multiple new plasma lines were identified with energies that correspond to known electronic transitions in He or Ne. The shifts of several of the lines are consistent with second-order diffraction in the UV region of the electromagnetic spectrum. This work complements published He-Ne laser emission line compilations and refines previously measured plasma peak shifts through use of a spectrometer with higher resolution. With an estimated uncertainty of $\sim$0.1 cm$^{-1}$, the data reported in the present work constitute the most complete and accurate compilation of He-Ne emission line wavelengths to date and should prove useful to those performing spectroscopy experiments that employ a He-Ne laser as a calibration or incident light source in the visible and ultraviolet spectral regions.
\bibliographystyle{apsrev4-1}
\bibliography{refs.bib}

\begin{thebibliography}{9}%
\makeatletter
\providecommand \@ifxundefined [1]{%
 \@ifx{#1\undefined}
}%
\providecommand \@ifnum [1]{%
 \ifnum #1\expandafter \@firstoftwo
 \else \expandafter \@secondoftwo
 \fi
}%
\providecommand \@ifx [1]{%
 \ifx #1\expandafter \@firstoftwo
 \else \expandafter \@secondoftwo
 \fi
}%
\providecommand \natexlab [1]{#1}%
\providecommand \enquote  [1]{``#1''}%
\providecommand \bibnamefont  [1]{#1}%
\providecommand \bibfnamefont [1]{#1}%
\providecommand \citenamefont [1]{#1}%
\providecommand \href@noop [0]{\@secondoftwo}%
\providecommand \href [0]{\begingroup \@sanitize@url \@href}%
\providecommand \@href[1]{\@@startlink{#1}\@@href}%
\providecommand \@@href[1]{\endgroup#1\@@endlink}%
\providecommand \@sanitize@url [0]{\catcode `\\12\catcode `\$12\catcode
  `\&12\catcode `\#12\catcode `\^12\catcode `\_12\catcode `\%12\relax}%
\providecommand \@@startlink[1]{}%
\providecommand \@@endlink[0]{}%
\providecommand \url  [0]{\begingroup\@sanitize@url \@url }%
\providecommand \@url [1]{\endgroup\@href {#1}{\urlprefix }}%
\providecommand \urlprefix  [0]{URL }%
\providecommand \Eprint [0]{\href }%
\providecommand \doibase [0]{http://dx.doi.org/}%
\providecommand \selectlanguage [0]{\@gobble}%
\providecommand \bibinfo  [0]{\@secondoftwo}%
\providecommand \bibfield  [0]{\@secondoftwo}%
\providecommand \translation [1]{[#1]}%
\providecommand \BibitemOpen [0]{}%
\providecommand \bibitemStop [0]{}%
\providecommand \bibitemNoStop [0]{.\EOS\space}%
\providecommand \EOS [0]{\spacefactor3000\relax}%
\providecommand \BibitemShut  [1]{\csname bibitem#1\endcsname}%
\let\auto@bib@innerbib\@empty
\bibitem [{\citenamefont {Loader}(1970)}]{Loader}%
  \BibitemOpen
  \bibfield  {author} {\bibinfo {author} {\bibfnamefont {J.}~\bibnamefont
  {Loader}},\ }\href {https://books.google.ca/books?id=bNPvAAAAMAAJ} {\emph
  {\bibinfo {title} {Basic Laser Raman Spectroscopy}}}\ (\bibinfo  {publisher}
  {Heyden},\ \bibinfo {year} {1970})\BibitemShut {NoStop}%
\bibitem [{\citenamefont {Craig}\ and\ \citenamefont {Levin}(1979)}]{Craig}%
  \BibitemOpen
  \bibfield  {author} {\bibinfo {author} {\bibfnamefont {N.~C.}\ \bibnamefont
  {Craig}}\ and\ \bibinfo {author} {\bibfnamefont {I.~W.}\ \bibnamefont
  {Levin}},\ }\href {https://opg.optica.org/as/abstract.cfm?URI=as-33-5-475}
  {\bibfield  {journal} {\bibinfo  {journal} {Appl. Spectrosc.}\ }\textbf
  {\bibinfo {volume} {33}},\ \bibinfo {pages} {475} (\bibinfo {year}
  {1979})}\BibitemShut {NoStop}%
\bibitem [{\citenamefont {Aktas}\ \emph {et~al.}(2011)\citenamefont {Aktas},
  \citenamefont {Truong}, \citenamefont {Otani}, \citenamefont {Balakrishnan},
  \citenamefont {Clouter}, \citenamefont {Kimura},\ and\ \citenamefont
  {Quirion}}]{Aktas_2012}%
  \BibitemOpen
  \bibfield  {author} {\bibinfo {author} {\bibfnamefont {O.}~\bibnamefont
  {Aktas}}, \bibinfo {author} {\bibfnamefont {K.~D.}\ \bibnamefont {Truong}},
  \bibinfo {author} {\bibfnamefont {T.}~\bibnamefont {Otani}}, \bibinfo
  {author} {\bibfnamefont {G.}~\bibnamefont {Balakrishnan}}, \bibinfo {author}
  {\bibfnamefont {M.~J.}\ \bibnamefont {Clouter}}, \bibinfo {author}
  {\bibfnamefont {T.}~\bibnamefont {Kimura}}, \ and\ \bibinfo {author}
  {\bibfnamefont {G.}~\bibnamefont {Quirion}},\ }\href {\doibase
  10.1088/0953-8984/24/3/036003} {\bibfield  {journal} {\bibinfo  {journal} {J.
  Condens. Matter Phys}\ }\textbf {\bibinfo {volume} {24}},\ \bibinfo {pages}
  {036003} (\bibinfo {year} {2011})}\BibitemShut {NoStop}%
\bibitem [{\citenamefont {Wojdyr}(2010)}]{fityk}%
  \BibitemOpen
  \bibfield  {author} {\bibinfo {author} {\bibfnamefont {M.}~\bibnamefont
  {Wojdyr}},\ }\href {\doibase 10.1107/S0021889810030499} {\bibfield  {journal}
  {\bibinfo  {journal} {J. Appl. Crystallogr.}\ }\textbf {\bibinfo {volume}
  {43}},\ \bibinfo {pages} {1126} (\bibinfo {year} {2010})}\BibitemShut
  {NoStop}%
\bibitem [{\citenamefont {Fuhr}\ and\ \citenamefont
  {Wiese}(2005)}]{fuhr2005nist}%
  \BibitemOpen
  \bibfield  {author} {\bibinfo {author} {\bibfnamefont {J.}~\bibnamefont
  {Fuhr}}\ and\ \bibinfo {author} {\bibfnamefont {W.}~\bibnamefont {Wiese}},\
  }\href@noop {} {\bibfield  {journal} {\bibinfo  {journal} {CRC Handb. Chem.
  Phys.}\ ,\ \bibinfo {pages} {128}} (\bibinfo {year} {2005})}\BibitemShut
  {NoStop}%
\bibitem [{\citenamefont {Ciddor}(1996)}]{Ciddor96}%
  \BibitemOpen
  \bibfield  {author} {\bibinfo {author} {\bibfnamefont {P.~E.}\ \bibnamefont
  {Ciddor}},\ }\href {\doibase 10.1364/AO.35.001566} {\bibfield  {journal}
  {\bibinfo  {journal} {Appl. Opt.}\ }\textbf {\bibinfo {volume} {35}},\
  \bibinfo {pages} {1566} (\bibinfo {year} {1996})}\BibitemShut {NoStop}%
\bibitem [{\citenamefont {Martin}(1960)}]{Martin}%
  \BibitemOpen
  \bibfield  {author} {\bibinfo {author} {\bibfnamefont {W.~C.}\ \bibnamefont
  {Martin}},\ }\href@noop {} {\bibfield  {journal} {\bibinfo  {journal}
  {Journal of Research of the National Bureau of Standards. Section A, Physics
  and Chemistry}\ }\textbf {\bibinfo {volume} {64}},\ \bibinfo {pages} {19}
  (\bibinfo {year} {1960})}\BibitemShut {NoStop}%
\bibitem [{\citenamefont {Wiese}\ \emph {et~al.}(1966)\citenamefont {Wiese},
  \citenamefont {Smith},\ and\ \citenamefont {Glennon}}]{GovPub}%
  \BibitemOpen
  \bibfield  {author} {\bibinfo {author} {\bibfnamefont {W.~L.}\ \bibnamefont
  {Wiese}}, \bibinfo {author} {\bibfnamefont {M.~W.}\ \bibnamefont {Smith}}, \
  and\ \bibinfo {author} {\bibfnamefont {B.}~\bibnamefont {Glennon}},\ }\href
  {https://apps.dtic.mil/sti/citations/AD0634145} {\emph {\bibinfo {title}
  {Atomic transition probabilities. Volume 1. Hydrogen through neon}}}\
  (\bibinfo  {publisher} {Washington, DC: National Bureau of Standards},\
  \bibinfo {year} {1966})\BibitemShut {NoStop}%
\bibitem [{\citenamefont {Herzberg}(1958)}]{Herzberg}%
  \BibitemOpen
  \bibfield  {author} {\bibinfo {author} {\bibfnamefont {G.}~\bibnamefont
  {Herzberg}},\ }\href@noop {} {\bibfield  {journal} {\bibinfo  {journal}
  {Proceedings of the Royal Society of London. Series A. Mathematical and
  Physical Sciences}\ }\textbf {\bibinfo {volume} {248}},\ \bibinfo {pages}
  {309} (\bibinfo {year} {1958})}\BibitemShut {NoStop}%
\end{thebibliography}%

\end{document}